\documentclass[aps,amssymb,pre,groupedaddress,twocolumn,showpacs]{revtex4-1}

\usepackage{graphicx}% Include figure files
\usepackage{dcolumn}% Align table columns on decimal point
\usepackage{bm}% bold math
\usepackage{amsfonts}
\usepackage{amsmath, amssymb}
\usepackage{color}
\usepackage{multirow}

\newcommand\qed{\phantom{\underline{y}}\hfill\hfill$\square$}

\newcommand\op[1]{\mathop{\rm #1}\nolimits}

\newcommand\R{{\mathbb R}}
\newcommand\E{{\mathbb E}}

\newcommand{\weg}[1]{}
\newcommand\ed{\stackrel{d}{=}}
\newcommand\dd{\stackrel{d}{\sim}}

\begin{document}

\title{Approximated maximum likelihood estimation in multifractal random walks}
\author{O. L{\o}vsletten}
\email[]{ola.lovsletten@uit.no}
\author{M. Rypdal}
\email[]{martin.rypdal@uit.no}
\affiliation{Department of Mathematics and Statistics, University of Troms{\o}, Norway.}


\begin{abstract}
We present an approximated maximum likelihood method for the multifractal random walk processes of [E. Bacry {\em et al.}, Phys. Rev. E {\bf 64}, 026103 (2001)]. The likelihood is computed using a Laplace approximation and a truncation in the dependency structure for the latent volatility. The procedure is implemented as a package in the R computer language. Its performance is tested on synthetic data and compared to an inference approach based on the generalized method of moments. The method is applied to estimate parameters for various financial stock indices.    
\end{abstract}

\maketitle

\section{Introduction}
Multifractal models were first introduced in the 1960s by the so-called ``Russian school'' in turbulence theory \citep{Obukhov1962,Kolmogorov1962}. In turbulence,  multifractality can be conceived as a weakening of the spatial selfsimilarity in the velocity field  implicitly assumed in Kolmogorov's 1941-theory \citep{Kolmogorov1941}. This generalization is called the Kolmogorov-Obukhov model and entails modeling the spatial variability of the energy dissipation rate as a random measure with certain multi-scaling properties. The Kolmogorov-Obukhov model is treated rigorously by Kahane \cite{Kahane1985} and  this construction is known as Gaussian multiplicative chaos.  

In recent years multifractal random processes and multifractal random measures have received increased attention and  are widely used in physics, geophysics and complex systems theory. Examples include  phenomena as diverse as internet traffic \cite{Riedi1999}, geomagnetic activity \cite{Rypdal2010,Rypdal2011} and rainfall patterns \cite{Pathirana2002}. In addition, multifractal processes provide natural models for the long-range volatility persistence observed in financial time series. This was first discovered by 
 Ghashghaie \cite{Ghashghaie1996} and Mandelbrot \cite{Mandelbrot1997}, and since the late 1990s  much work has been done on multifractal modeling of financial markets  \citep{DiMatteo2007,Calvet2001}. 
Logarithmic returns of assets are modeled as $x_t = X(t+\Delta t)-X(t)$, where $X(t)$ are continuous-time processes with stationary increments and multifractal scaling. The latter means that the moments of $X(t)$ are power-laws as functions of time;
\begin{equation}
\E|X(t)|^q \sim  t^{\zeta(q)},
\label{eq1}
\end{equation}
either in some interval $t \in (0,R)$ or asymptotically as $t \to 0$. The scaling function $\zeta(q)$ is linear for self-similar processes, but may in general be concave. Processes satisfying equation (\ref{eq1}) with strictly concave scaling functions are generally referred to as  multifractal.

Two well-known ``stylized facts'' of financial time series are that log-returns are uncorrelated and non-Gaussian. Based on this, Mandelbrot \cite{Mandelbrot1963} deduced that if prices are described as selfsimilar processes, then these processes must be so-called L{\'e}vy flights, i.e. $\alpha$-stable L{\'e}vy processes with $\alpha<2$. However, if one allows non-linear scaling functions, then one can maintain uncorrelated log-returns by simply imposing the condition $\zeta(2)=1$. The concave shape of $\zeta(q)$ implies that the variables $X(t)$ are increasingly leptokurtic with decreasing $t$, and consequently non-Gaussian. Moreover, as opposed to L{\'e}vy flights, multifractal processes have strongly dependent increments and can therefore describe a third ``stylized fact'' of financial time series, namely volatility clustering. 

Notwithstanding that multifractal processes provide accurate and parsimonious descriptions of temporal financial fluctuations, the models are rarely implemented for forecasting and risk-analysis in financial institutions. This is partially due to a lack of accurate, stable and efficient inference methods for multifractal processes.        
Parameter estimation has so far mostly been made using various moment-based estimators, such as the generalized method of moments (GMM). Alternatively, one can fit the estimated scaling functions to theoretical expressions of $\zeta(q)$. However, as pointed out in e.g. \cite{Lux2003} and \cite{Chapman2005}, the standard estimators of scaling exponents have large mean square errors for time series of length comparable to those typically available in econometrics. 

An exception to the statements above is the  Markov-Switching Multifractal (MSM) model \citep{Calvet2001} where maximum likelihood estimation is feasible. In discrete time MSM implies that the increments $x_t$ are described by a stochastic volatility model on the form 
\begin{equation} \label{MSM}
x_t = \sigma\,\sqrt{M_t}\, \varepsilon_t\,.
\end{equation}
Here $\varepsilon_t \dd \mathcal{N}(0,1)$ are independent variables and the volatility is a product on the form
$$
M_t=M_{t,1} M_{t,2} \cdots M_{t,K}\,,
$$
where (for each time step $t$) $M_{t,k}$ are updated from a distribution $M$ with a probability $\gamma_k=1-(1-\gamma_1)^{b^{k-1}}$.   
In this model however, maximum likelihood estimation is only possible in the case where $M$ is defined on a discrete state space, and there is a limitation on the magnitude of $K$ which should not exceed $\approx 10$ \citep{Lux2008}. These restrictions not only limit flexibility with respect to the distribution of returns, but also the possible range in the volatility dependency.

This paper concerns parametric inference for the multifractal random walk (MRW) introduced by Bacry et al.  \cite{Bacry2001}. The increment process $x_t=X(t+\Delta t)-X(t)$ is  still a discrete-time stochastic process  described by equation (\ref{MSM}), 
but now the volatility is modeled as $M_t=c\,e^{h_t}$, where $h_t$  is a stationary and centered Gaussian process with the  co-variance structure
\begin{equation} \label{defmod2}
\text{Cov}(h_t,h_s)=\lambda^2\log^+\frac{T}{(|t-s|+1)\Delta t}\,,
\end{equation}
where $\log^+ a \stackrel{\op{def}}{=} \max\{\log a, 0\}$. 
Here $T>0$ is called the correlation range \footnote{In turbulence $T$ corresponds to the integral scale.} and $\lambda>0$ is called the intermittency parameter. The constant $c$ ensures normalization and is chosen so that $1/c=\E[e^{h_t}]$. We denote $R=T/\Delta t$.

Let $\theta=(\lambda,\sigma,R)$ denote the parameter vector and $y=(y_1,\dots,y_n) \in \R^n$ a fixed time series. The main result of this paper is the development of  a method for efficiently computing approximations to the likelihood function 
$$
{\mathcal L}(\theta|y) = p_x(y|\theta)\,,
$$
where $p_x(\cdot|\theta)$ is the probability density function of a random vector $x=(x_1,\dots,x_n)$ produced by the MRW model with parameters $\theta$. Using the likelihood function, parameters can be determined by means of the maximum likelihood (ML) estimator: 
$$
\hat{\theta} = \op{argmax}_{\theta} {\mathcal L}(\theta|y)\,.
$$

Our method exploits that the discrete MRW model has a construction  similar to simple volatility (SV) models. The distinguishing feature is that the processes $h_t$  are autoregressive in SV models. By truncating 
the dependency structure in the logarithmic volatility $h_t$, the computation of the likelihood function is mapped on to a similar problem for SV models, and hence existing techniques for further approximations are available.   

To our knowledge the present paper is the first to present results on ML estimation for multifractal models with continuous state spaces for the volatility. Such estimates may be of great practical importance, since accurate parameter estimation is essential for volatility forecasts and  risk estimates. In the MRW model this degree of accuracy is particularly important for the intermittency parameter $\lambda$ which determines the peakedness of the return distributions on all time scales. 
In applications other than finance, accurate estimates of $\lambda$ can be used as supplements to the empirical scaling functions, and thereby the ML estimator can provide a tool for quantifying multifractality in data.

The paper is organized as follows: In section \ref{model} we briefly explain the construction of the continuous-time process  $X(t)$ for which the model given by equations (\ref{MSM}) and (\ref{defmod2}) is a discretization. 
There exists a large class of multifractal processes which are related to a construction known as infinitely divisible cascades (IDC).  
In general the random walk models associated with IDC processes have logartithmic volatility with infinitely divisible distributions, and the MRW model considered in this paper is a discrete approximation to the 
random walk model obtained in the special case when the logarithmic volatility is Gaussian.

Section \ref{method} contains the procedure for approximated ML estimation in the MRW model. In section \ref{montecarlo} we test the estimator by first applying it to various stock market indices, and then by running a small Monte Carlo study. The results are compared with the GMM method used in \citep{Bacry2008}. 

We finally remark that the methods presented in this paper have been implemented in a package for the R statistical software \citep{R}. 
This package is available online \cite{code}.

\section{Motivation of the model} \label{model}
There exist several popular models for multifractal stochastic processes with uncorrelated increments. All of these models can be written either on the form $X(t) = B(A(t))$, where $B(t)$ is a Brownian motion and $A(t)=m([0,t])$ is the distribution function of a multifractal random measure $m$ on the time axis, or as 
$$
X(t) = \lim_{r \to 0} \int_0^t \sqrt{A_r(t')}\,dB(t')\,,
$$
where $A_r(t) \to A(t)$ as $r \to 0$. The meaning of $A_r(t)$ is discussed below. These two types of constructions are equivalent as long as $B(t)$ is a Brownian motion. (This is not the case for fractional Brownian motions with $H \neq 1/2$.)

The differences between the various multifractal models are then related to the construction of the random measure $m$. The log-normal MRW model is on one hand based on a particular construction of $m$ known as multiplicative chaos, and on the other hand it can be seen as a special case of the more general IDC constructions.  

In multiplicative chaos, which was first developed rigorously in \cite{Kahane1985}, one considers a sequence $m_n$ of measures defined via random densities on the form 
$$
dm_n (t) = c_n\,e^{h_n(t)}\,dt\,,
$$ 
where $1/c_n=\E[e^{h_n(t)}]$, and $h_n(t)$ are centered Gaussian processes with co-variance structures $g_n(t,s)=\text{Cov}(h_n(t),h_n(s))$ that converge to some expression
$g(t,s)$ in the limit $n \to 0$. Kahane \cite{Kahane1985}  showed that if $g$ is $\sigma$-positive, meaning that there are positive and positive definite functions $K_m(t,s)$ such that  
$$
g_n(t,s) = \sum_{m = 1}^n K_m(t,s)\,,
$$
then the sequence $m_n$ converges weakly to a Borel measure $m$ which depends only on the function $g(t,s)$. One can therefore informally think of $m$ as being on the form $dm (t) = c\,e^{h(t)}\,dt$ where $1/c=\E[e^{h(t)}]$ and $h(t)$ is a ``Gaussian'' process with co-variance $g(t,s)$. Then, if one makes the choice 
\begin{equation} \label{covar}
\gamma(t,s) = \lambda^2 \log^+ \frac{R}{|t-s|}\,,
\end{equation}
one easily obtains the relation  $h(at)\ed h(t)+\Omega(c)$, where $\Omega(a)$ are independent of $h(t)$ and distributed according to 
$\Omega(a) \dd \mathcal{N}(0,-\lambda^2\log a)$. It follows that we for $t<R$ and $0<a<1$ have the scaling relation  
\begin{equation} \label{scaling1}
m\left([0,at]\right)\ed M(a)m([0,t]) \,,
\end{equation}
with $\log M(a) \dd  \mathcal{N}\left(\left(1 +\lambda^2/2\right)\log a,-\lambda^2\log a\right)$.
See proposition 3.3 in \citep{Robert2010} for a rigorous proof of (\ref{scaling1}), and see example 2.3 of the same paper for a verification that the function $g(t,s)$ in equation (\ref{covar}) is $\sigma$-positive. 
By using the well-known formula for the $q$-th moments of log-normal variables together with 
equation (\ref{scaling1}), we easily verify the multifractality of the process $A(t)=m([0,t])$: Denote $C_q = \E |m([0,1])|^q$ and observe that
$$
\E |m([0,t])|^q = C_q\,\E M(t)^q = C_q t^{\zeta_A(q)}\,,
$$  
where $\zeta_A(q)=q\left(1 +\lambda^2/2\right)-\lambda^2q^2/2$. Since a Brownian motion is self-similar with $H=1/2$ the scaling function of $X(t)$ is given by $\zeta(q)=\zeta_A(q/2)$. 

Alternatively the model defined by equations (\ref{MSM}) and (\ref{defmod2}) can be motivated by considering the more general class of IDC models. Here we briefly mention the main ideas and results in this theory, and we refer to \cite{Bacry2003,Muzy2002} for details.
At the base of this construction is an object called an independently scattered infinitely divisible random measure $P(dt,dr)$ defined on the halfplane $S^+=\{(t,r) \in \R^2\,|\,r \geq 0\}$. The defining properties of the random measure $P$ are: (1)
for any measurable set ${\mathcal A} \subset S^+$, the random variable $P({\mathcal A})$ is infinitely divisible with characteristic function 
$$
\varphi_{P(A)}(q) = e^{\psi(q) \mu({\mathcal A})}\,, 
$$  
where $\mu(dt,dr)=r^{-2}\,dt dr$. (2) for any finite sequence $\mathcal{A}_k \subset S^+$ of disjoint and measurable sets, the corresponding random variables $P({\mathcal A}_k)$ are independent. 
If we assume that $\psi'(0)=0$, the random measure $P$ induces a family of centered and stationary stochastic processes through the equation 
$$
h_r(t) = P\big{(} \mathcal{A}(r,t)\big{)}\,,
$$ 
where $\mathcal{A}(r,t)$ are cone-like domains defined by 
$$
\mathcal{A}(r,t) = \big{\{} (t',r') \in S^+ \,|\, r' \geq r,\, |t'-t|  \leq f(r')/2 \big{\}} \,,
$$
with $f(r)=r$ for $r \leq R$ and $f(r)=R$ for $r > R$. 
The time correlations in the processes $h_r(t)$ are characterized by the functions 
\begin{equation*}
\begin{array}{lll}
\rho_r(t) &=& \displaystyle \mu ( \mathcal{A}(0,r) \cap \mathcal{A}(t,r))  \\~ \\ &=& \displaystyle 
\begin{cases} 
\log \frac{R}{r}+1-\frac{t}{r} &,\,\, t < r \\ 
\log^+ \frac{R}{t} &,\,\, t\geq r
\end{cases}\,.
\end{array}
\end{equation*}
In fact, the co-variance of $h_r(t)$ is given by 
$$
\op{Cov}(h_r(t),h_r(s)) = \lambda^2 \rho_r(|t-s|)\,,
$$ 
where $\lambda^2 = -\psi''(0)$.

Random measures are defined by $dm_r(t) = c_r e^{h_r(t)}\,dt$, where $1/c_r = \E[e^{h_r(t)}]$. The
corresponding distribution functions are $A_r(t)=m_r([0,t])$ and corresponding random walks are 
$$
X_r(t) = \int_0^t \sqrt{A_r(t')}\,dB(t')\,.
$$
By using the relation $\rho_{ar}(at) = -\log a + \rho_r(t)$ one can show that 
$$
h_{ar}(at) \ed h_r(t)+\Omega(a)\,,
$$ 
for $a \in (0,1)$ and $t \leq R$, 
where $\Omega(a)$ are independent of $h_r(t)$ and have characteristic functions $\varphi_{\Omega(a)}(q) = e^{-\psi(q) \log a}$. 
Consequently the limit process $X(t) = \lim_{r \to 0} X_r(t)$ has scaling function 
$$
\zeta(q) = \big{(} 1+\psi(-i) \big{)}\,q/2-\psi(-iq/2)\,.
$$
In the case that $h_r(t)$ are Gaussian, i.e. $\psi(q) = -\lambda^2 q^2/2$, the co-variance is on the form 
$$
\op{Cov}(h_r(t),h_r(s)) = \lambda^2 \log^+ \frac{R}{|t-s|}
$$
for $|t-s|>r$, and hence it can be approximated by the process defined by equation (\ref{defmod2}). 
In this case the scaling function is 
$$
\zeta(q) = \big{(} 1+\lambda^2/2 \big{)}\,q/2-\lambda^2 q^2/8\,.
$$
We note that for $\lambda=0$ the process $X(t)$ is reduced to a Brownian motion and $\zeta(q)=q/2$.

We point out that this paper presents a ML estimator for the discrete-time process $x_t$ defined by equations (\ref{MSM}) and (\ref{defmod2}). This is sufficient for the purpose of modeling and forecasting volatility in financial time series, since the discrete-time MRW model is directly comparable to GARCH-type models. In other applications, such as modeling the velocity field in turbulence, one is interested in the continuous-time process $X(t)$. Since $x_t$ is an approximation to the continuous-time process $X(t)$, our method can also be interpreted as an estimator for this process. In this case one must be aware that the increment process $X(t+\Delta t)-X(t)$ is not proportional (in law) to $e^{h_{\Delta t}(t)} \varepsilon_{t}$, and that this is only an approximation in the limit $\lambda^2 \ll 1$. See appendix A.1. in \citep{Bacry2008}. In the case of strong intermittency, the estimator for the continuous-time process $X(t)$ may therefore depend significantly on the time-scale $\Delta t$ for which the data is sampled. An analysis of how our method performs as an estimator for $X(t)$ will require extensive Monte Carlo simulations (with varying $\lambda$ and $\Delta t$), and this is beyond the scope of this paper.

We also remark that it in some applications is relevant to estimate the parameters of the measure $dm(t)$, for instance when modeling the energy dissipation fields in turbulence. In the discrete-time approximation this corresponds to the process $e^{h_t}$, where $h_t$ is described by equation (\ref{defmod2}). Since $h_t$ is Gaussian, this problem is much easier than the one considered in this paper. The ML estimator for $e^{h_t}$ can be constructed using standard methods \cite{McLeod2007} and no approximations are required.

\section{Approximated maximum likelihood} \label{method}
In this section we explain our method of approximated maximum likelihood estimation. Let $x_t$ and $h_t$ be the processes defined by (\ref{MSM}) and (\ref{defmod2}). Denote $x=(x_1,\dots,x_n)$ and  $h=(h_1,\dots,h_n)$. The first step is to write 
\begin{equation} \label{eq11}
p_x(x)=\int_{\mathbb{R}^n} p_{x,h}(x,h)\,dh=\int_{\mathbb{R}^n} p_{x|h}(x|h)p_h(h)dh.
\end{equation}
The first factor $p_{x|h}(x|h)$ in the integrand is computed by noting that, when conditioned on $h$, the variables $x_1,\dots,x_n$ are independent and Gaussian. In fact,
\begin{equation} 
\begin{array}{lll}
\displaystyle \log p_{x|h}(x|h) &=& \displaystyle \sum_{t=1}^n \log p_{x_t|h_t}(x_t|h_t) \\
~ &=& \displaystyle -n \log \sqrt{2 \pi c}\, \sigma  \\
~ &+& \displaystyle   \sum_{t=1}^n \Big{(}-\frac{h_t}{2} - \frac{x_t^2}{2\, \sigma^2 c\, e^{h_t}} \Big{)}\,.
\end{array} 
\label{factor1}
\end{equation}
For the second factor $p_h(h)$ we use that $h_t$ is a centered Gaussian process with a specified co-variance structure $\text{Cov}(h_t,h_s)=\gamma(|t-s|)$. First we decompose the density into one-dimensional marginals;
\begin{equation} \label{marginals}
	p_h(h) = p_{h_1}(h_1)\prod_{t=2}^n p_{h_t|h_{1:t-1}}(h_t|h_{1:t-1})\,,
\end{equation} 
where we have used the notation 
$$
h_{n:m} = 
\begin{cases}
(h_n,h_{n+1},\dots,h_m) & \text{ for }m\geq n \\
(h_n,h_{n-1},\dots,h_m) & \text{ for }n>m 
\end{cases}\,.
$$
Denote by $\Gamma_t$ the co-variance matrix of the vector $h_{1:t}$, and let $\gamma_{1:t} = (\gamma(1),\dots,\gamma(t))$. The co-variance matrix can be written on the  block form: 
$$
\Gamma_t = 
\begin{pmatrix}
\gamma(0) &  \gamma_{1:t-1} \\ \gamma_{1:t-1}^T &\Gamma_{t-1} 
\end{pmatrix}\,.
$$ 
By performing standard computations of conditional marginals in multivariable Gaussian distributions we deduce that $h_t|h_{1:t-1}$ is a Gaussian with mean  
$$
m_t =  \gamma_{1:t-1} \Gamma_{t-1}^{-1} h_{(t-1):1}^T
$$ 
and variance 
$$
S_t^2 = \gamma(0) -  \gamma_{1:t-1} \Gamma_{t-1}^{-1}  \gamma_{1:t-1}^T\,. 
$$
As usual it is convenient to introduce vectors  $\phi^{(t)}$ defined by 
$\phi^{(t)} \Gamma_{t} =\gamma_{1:t}$. This
allows us to write the mean as $m_t = \phi^{(t-1)} h_{(t-1):1}^T$ and the variance as 
$S_t^2 =  \gamma(0) -  \phi^{(t-1)}  \gamma_{1:t-1}^T$. Then from equation (\ref{marginals}) we have 
\begin{equation} 
\begin{array}{lll}
\log p_h(h)  &=& \displaystyle - n \log \sqrt{2 \pi} - \sum_{t=1}^n \log S_t \\  
~\\
&-& \displaystyle \sum_{t=1}^n \frac{(h_t-\phi^{(t-1)} h_{(t-1):1}^T)^2}{2 S_t^2} \,. 
\end{array}
\label{factor2}
\end{equation}
Combining equation (\ref{factor2}) with equation (\ref{factor1}) we get an expression for the full likelihood: 
\begin{equation} 
\begin{array}{lll}
\log p_{x,h}(x,h) &=& \displaystyle -n \log (2 \pi\,\sqrt{c}\,\sigma) \\
 &+& \displaystyle \sum_{t=1}^n \Big{(}-\frac{h_t}{2}  - \frac{x_t^2}{2 \sigma^2 \,c\,e^{h_t}} \Big{)}  \\ 
&-& \displaystyle \sum_{t=1}^n \log S_t   \\
 &-& \displaystyle \sum_{t=1}^n \frac{(h_t- \phi^{(t-1)} h_{(t-1):1}^T)^2}{2 S_t^2} \,.
\end{array}
\end{equation}
We keep in mind that $c$ depends on $R$ and $\lambda$ through the relation 
$1/c= \E[e^{h_t}]=R^{\lambda^2/2}$. \\

\noindent {\bf Approximation 1:} By comparing co-variances the process $h_t$ can be written as   
\begin{equation} \label{AR}
h_t = \phi^{(t-1)}_{1} h_{t-1} + \dots +  \phi^{(t-1)}_{t-1} h_{1} + w_t\,,
\end{equation}
where $w_t$ are independent Gaussian variables with zero mean and variances equal to $S_t^2$. As approximations to $h_t$ we can consider processes obtained by truncating the sum in equation (\ref{AR}). We fix a parameter $\tau \in {\mathbb N}$ , and for $t>\tau$ we replace equation (\ref{AR}) with
\begin{equation} \label{ARtrunc}
h_t = \phi^{(\tau)}_1 h_{t-1} + \dots +  \phi^{(\tau)}_{\tau} h_{t-\tau} + w^{(\tau)}_t\,,
\end{equation}
where $w_t^{(\tau)}$ are independent Gaussian variables with zero mean and variances equal to $S_{\tau+1}^2$. Note that $h_t| h_{t-1:t-\tau}^T$ in equation (\ref{AR}) has the same distribution as obtained from (\ref{ARtrunc}), namely a Gaussian with mean $m_t = \phi^{(\tau)} h_{(t-1):t-\tau}^T$ and variance $S_{\tau+1}^2$. In effect we have approximated the distribution of $h_t| h_{t-1:1}$, by truncating the dependency after a lag $\tau$. As a result of this approximation equation (\ref{factor2}) becomes
\begin{equation} \label{factor3}
\begin{array}{lll}
\log p_h(h)  &=& \displaystyle - n \log \sqrt{2 \pi} \\
&-& \displaystyle \sum_{t=1}^\tau \log S_t  - (n-\tau) \log S_{\tau+1}  \\  
&-& \displaystyle \sum_{t=1}^\tau \frac{(h_t- \phi^{(t-1)} h_{(t-1):1}^T)^2}{2 S_t^2} \\
&+& \displaystyle \sum_{t=\tau+1}^n \frac{(h_t- \phi^{(\tau)} h_{t-1:t-\tau}^T)^2}{2 S_{\tau+1}^2} \,. 
\end{array}
\end{equation}
In order to compute the expression in equation (\ref{factor3}) we need to solve the equations 
$$
\phi^{(t)} \Gamma_{t} = \gamma_{1:t} \,,\,\,t=1,\dots,\tau\,.
$$
This is done efficiently using the Durbin-Levinson algorithm \citep{Levinson1946, Trench1964}. We remark that for $\tau=n$ the expression in equation (\ref{factor3}) is exact. \\

\noindent {\bf Approximation 2:} The second approximation is the so-called Laplace's method, which is frequently used for approximation of likelihoods in SV models, see e.g. \cite{Skaug2009,Aas2011}. We write equation (\ref{eq11}) on the form 
\begin{equation} \label{laplace1}
p_x(x) = \int_{\mathbb{R}^n} e^{n f_x({h})} dh\,,
\end{equation}
where 
\begin{equation} \label{f}
\begin{array}{lll}
f_x(h) &=& \displaystyle \frac{1}{n} \log p_{x,h}(x,h) \\ &=& \displaystyle \frac{1}{n} \sum_{t=1}^n \log p_{x_t|h_t}(x_t) \\ &+& \displaystyle \frac{1}{n} \sum_{t=1}^n \log p_{h_t|h_{1:t-1}} (h_t)\,.
\end{array}
\end{equation}

Laplace's method is to assume that $f_{x}(h)$ has a global maximum in $\mathbb{R}^n$, which we denote by $h^*$. When $n$ is large the contribution to the integral in equation (\ref{laplace1}) is concentrated around $h^*$, and therefore we make a second order Taylor approximation to $f_{x} (h)$ around this point. Since $h^*$ is also a local maximum we have
$$
f_{x}(h) \approx \frac1n \log p_{x,h}(x,h^*) + \frac{1}{2n} (h-h^*) \,\Omega_x\, (h-h^*)^T\,,
$$
where 
$$
\Omega_x=\frac{\partial^2  \log p_{x,h}(x,h^*)}{\partial h\,\partial h^T}
$$
is the Hessian matrix of $f_x (h)$ at the point $h^*$. The approximation now reads 
\begin{equation*}
\begin{array}{lll}
p_x(x) &\approx& \displaystyle e^{f_x(h^*)}\int_{\mathbb{R}^n} e^{\frac{1}{2} (h-h^*) \,\Omega_x\, (h-h^*)^T} dh \\ &=& \displaystyle (2 \pi)^{n/2}\,|\det \Omega_x|^{-1/2}\,p_{x,h}(x,h^*)\,.
\end{array}
\end{equation*}
The maximum $h^*$ is found by computing the partial derivatives of $f_x(h)$ with respect to $h$, setting them equal to zero and solving the corresponding system of equations numerically using the algorithm DF-SANE \citep{Raydan2006}, which is implemented in R\, package ``BB'' \cite{BB}. The matrix $\Omega_x$ is obtained using analytical expressions for the second derivatives. This matrix is band-diagonal with bandwidth equal to the truncation parameter $\tau$, and in the R\, software such matrices are efficiently stored and manipulated  using the package ``Matrix'' \citep{Matrix}.

%%
%%
%**********************************
\noindent
\begin{figure}[t]
\begin{center}
\includegraphics[width=8.4cm]{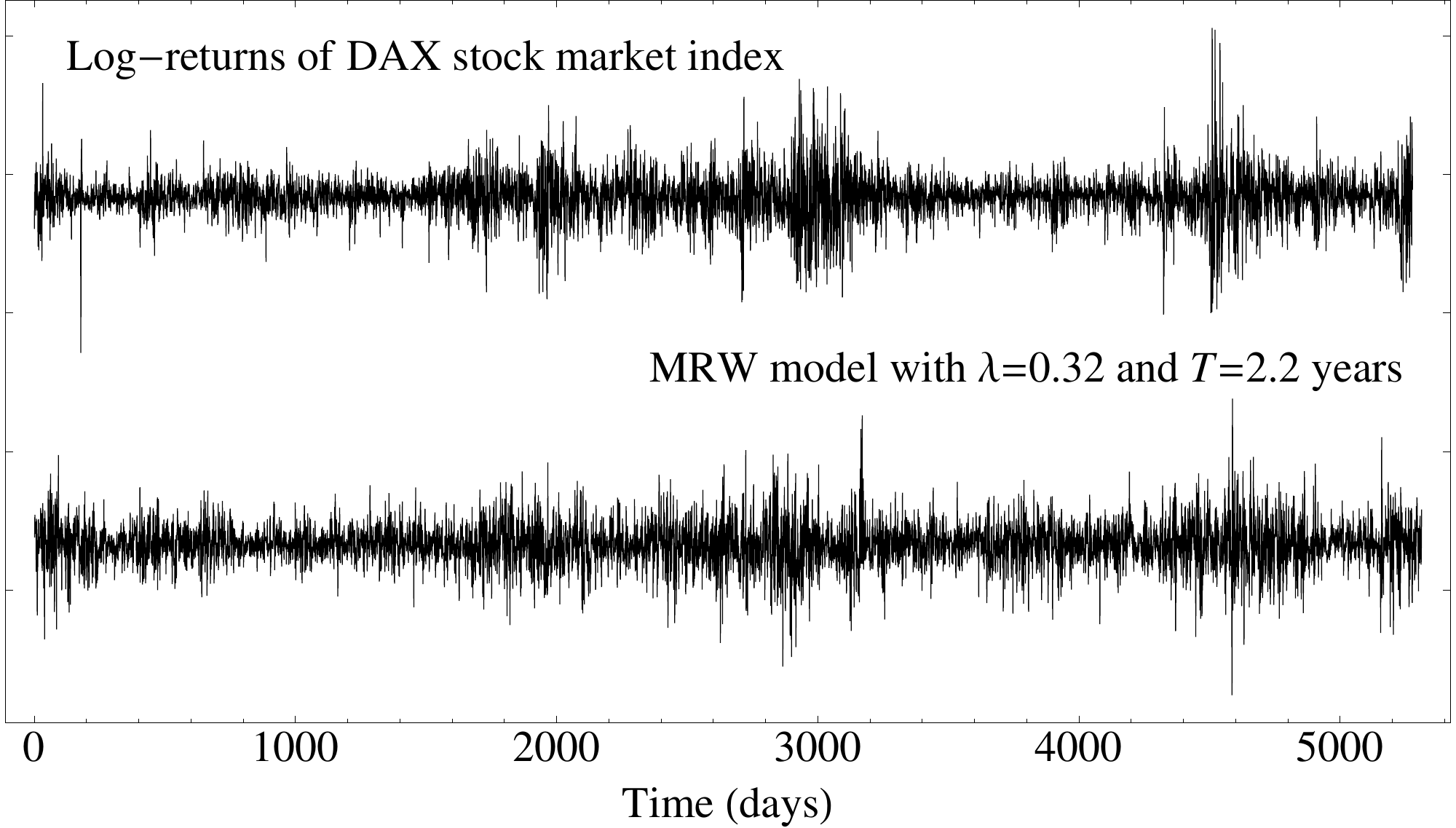}
\caption{The top figure shows the daily log-returns of the German DAX index for the time period 1990/11/26--2011/11/25. The standard deviation of the data is normalized to unity. For $\tau=500$ the ML estimates are $\lambda=0.32$ and $T=2.2$ years. The lower figure shows a simulation of the MRW model $x_t$ with the estimated parameters.} \label{Fig1}
\end{center}
\end{figure}

 \begin{table}[h!]  
 \caption{Estimated parameters for the log-returns of various stock market indices. Prior to the analysis the sample standard deviation of each data set is normalized to unity. All ML estimates are run with $\tau = 500$ and the GMM estimates are performed with a maximum time lag $t_{\op{max}} =500$ days in the auto-correlation function of $m_t=\log x_t^2$. The analyzed data is retrieved from \cite{yahoo}.} \label{tab1}
\begin{center}
\begin{tabular}{|lc||c|c||c|c|c||} \cline{3-6} 
\multicolumn{2}{c||}{~}    & \multicolumn{2}{|c||}{ML} &  \multicolumn{2}{|c|}{GMM}  \\ \cline{1-6}
Index   & (time period)  & $\lambda$                              & $T$ (years)            & $\lambda$ & $T$ (years)                          \\ \hline \hline
CAC 40 &   {\small (1990--2011)}          &  $0.29$                                 & $2.8$                          &         $0.36$                & $2.5$                              \\ \hline
S\&P 500 &   {\small (1950--2011)}        & $0.32$                                & $12.2$                              & $0.36$                & $10.2$                             \\ \hline
DAX &   {\small (1990--2011)}                & $0.32$                              & $3.3$                           & $0.44$               &  $3.1$                             \\ \hline
Nikkei 225 &   {\small (1984--2011)}       & $0.36$                              & $1.4$                                & $0.40$              &  $3.0$                            \\ \hline
Hang Seng &   {\small (1986--2011)}                 & $0.37$                             & $2.8$                              & $0.44$              &  $2.5$                           \\ \hline
FTSE 100 &   {\small (1984--2011)}       & $0.28$                             & $4.2$                              & $0.36$              &  $2.9$                           \\ \hline
\end{tabular}
\end{center}
\end{table}

\begin{table}[h!]  
\caption{The results of a Monte Carlo study of the ML and GMM estimators. The parameters in the simulations are $\lambda=0.35$ and $R=2000$ (i.e. $\log R=7.6$). In the GMM estimator we have used a maximum time lag $t_{\op{max}} =500$ days in the auto-correlation function of $m_t=\log x_t^2$. The reported values are the mean estimates together with the standard deviations (in brackets).} \label{tab2}
\begin{center}
\begin{tabular}{||c|c||c|c|c||c|c|c||} \cline{3-8} 
\multicolumn{2}{c||}{~} &  \multicolumn{3}{c||}{ML} & \multicolumn{3}{c||}{GMM} \\ \hline 
$n$                              & \small{$\tau$} & \small{$\lambda$} & \small{$\log R$}           & \small{$\sigma$}      & \small{$\lambda$} & \small{$\log R$} & \small{$\sigma$}  \\ \hline \hline
\multirow{6}{*}{2500}   & \multirow{2}{*}{10}   &  $0.31$                  & $6.87$                & $0.97$                  & ~~~                 & ~~                  & ~~    \\ 
                                    &                                 &   \small{$(0.03)$}             &  \small{$(3.41)$}              &  \small{$(0.19)$}              & ~~~                  & ~~                 & ~~   \\ \cline{2-5}  
			            & \multirow{2}{*}{50}   &   $0.34$                 & $6.47$                & $0.97$                 & $0.34$                  & $6.11$                & $0.97$     \\ 
                                    &                                  &  \small{$(0.03)$}  & \small{$(1.73)$}  & \small{$(0.19)$}   & \small{$(0.08)$}   & \small{$(0.76)$}  & \small{$(0.19)$}   \\ \cline{2-5} 
			            & \multirow{2}{*}{100} &  $0.34$                 & $6.35$                 & $0.97$                & ~~                   & ~~                  & ~~     \\ 
                                   &                                   &   \small{$(0.03)$}             &    \small{$(1.67)$}            &  \small{$(0.19)$}           & ~~                  & ~~~                 & ~~     \\ \hline \hline

\multirow{6}{*}{5000}   & \multirow{2}{*}{10}   &  $0.30$                  & $5.58$               & $0.98$                & ~~~                 & ~~                  & ~~    \\ 
                                    &                                 &   \small{$(0.03)$}               &  \small{$(2.18)$}              &  \small{$(0.14)$}              & ~~~                  & ~~                 & ~~   \\ \cline{2-5}  
			            & \multirow{2}{*}{50}   &  $0.34$                 & $7.02$                & $0.98$                 & $0.35$                  & $6.69$                & $0.981$     \\ 
                                    &                                  &  \small{$(0.02)$}  & \small{$(1.44)$}  & \small{$(0.14)$}   & \small{$(0.05)$}   & \small{$(0.96)$}  & \small{$(0.15)$}   \\ \cline{2-5} 
			            & \multirow{2}{*}{100} &  $0.34$                 & $6.87$                 & $0.97$                 & ~~                   & ~~                  & ~~     \\ 
                                   &                                   &   \small{$(0.02)$}              &    \small{$(1.31)$}             &  \small{$(0.14)$}            & ~~                  & ~~~                 & ~~     \\ \hline \hline
                                   
\multirow{6}{*}{10000}   & \multirow{2}{*}{10}   &  $0.30$                  & $9.10$                & $0.98$                  & ~~~                 & ~~                  & ~~    \\ 
                                    &                                 &   \small{$(0.02)$}               & \small{$(1.80)$}             &  \small{$(0.10)$}             & ~~~                  & ~~                 & ~~   \\ \cline{2-5}  
			            & \multirow{2}{*}{50}   &  $0.34$                 & $7.37$                & $0.98$                 & $0.35$                  & $7.11$                & $0.98$     \\ 
                                    &                                  &  \small{$(0.01)$}  & \small{$(1.24)$}  & \small{$(0.10)$}   & \small{$(0.04)$}   & \small{$(0.92)$}  & \small{$(0.10)$}   \\ \cline{2-5} 
			            & \multirow{2}{*}{100} &  $0.34$                 & $7.21$                 & $0.98$                 & ~~                   & ~~                  & ~~     \\ 
                                   &                                   &   \small{$(0.01)$}              &    \small{$(1.16)$}            &  \small{$(0.10)$}             & ~~                  & ~~~                 & ~~     \\ \hline \hline

\end{tabular}
\end{center}
\end{table}

%**********************************
\noindent
\begin{figure*}[t]
\begin{center}
\includegraphics[width=16.0cm]{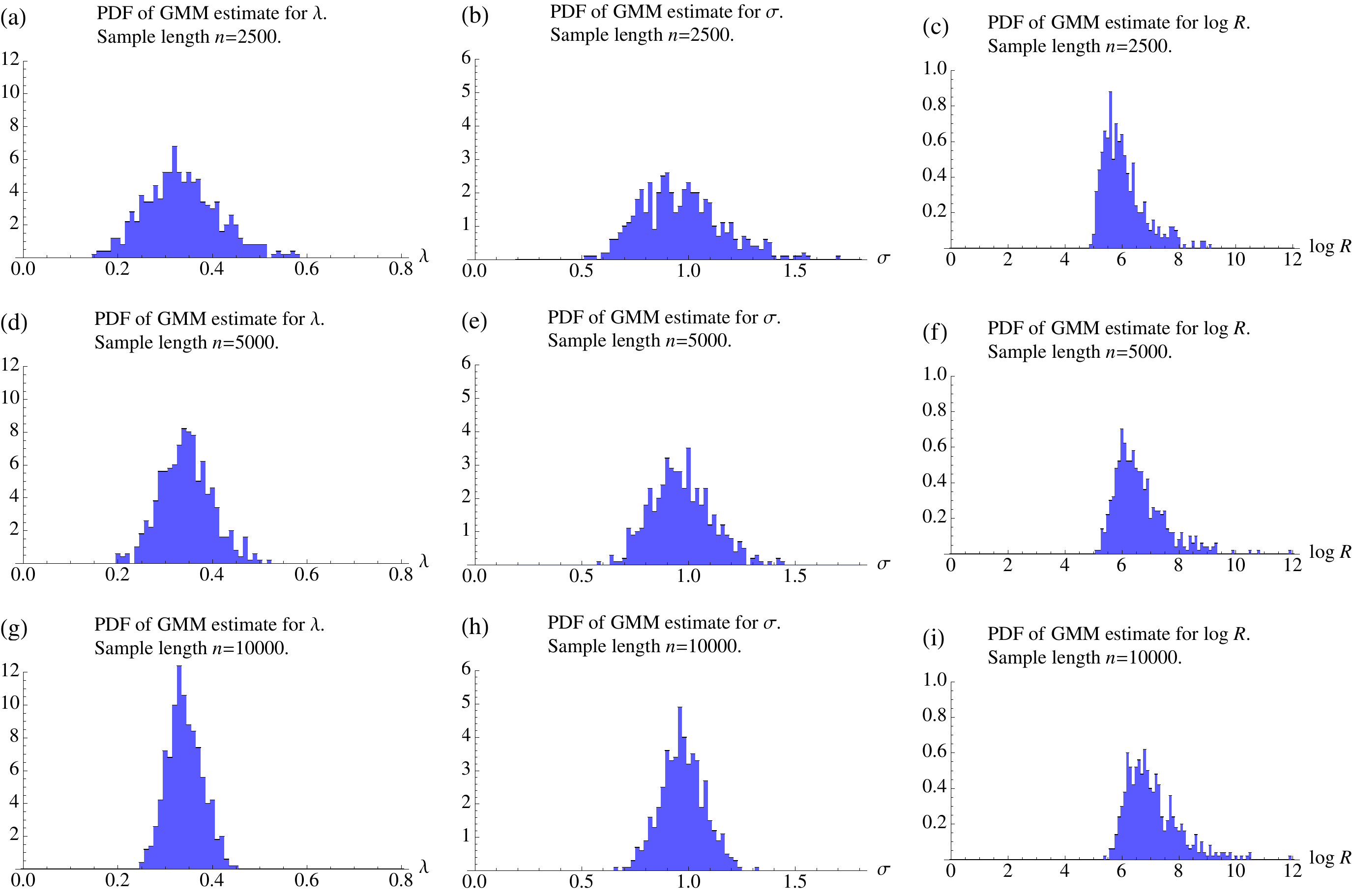}
\caption{The results of the Monte Carlo study for the GMM estimator explained in section \ref{montecarlo}. The figures show the estimated probability density functions for the estimators based on 500 realizations of the process. The parameters are $\lambda=0.35$, $\sigma=1$ and $R=2000$ (i.e. $\log R = 7.6$). In figures (a-c) the sample lengths are $n=2500$, in figures (d-f) the sample lengths are $n=5000$ and in figures (g-i) the sample lengths are $n=10000$. The means and standard deviations of the estimators are reported in table \ref{tab2}.} \label{Fig2}
\end{center}
\end{figure*}
%**********************************

%**********************************
\noindent
\begin{figure*}[t]
\begin{center}
\includegraphics[width=16.0cm]{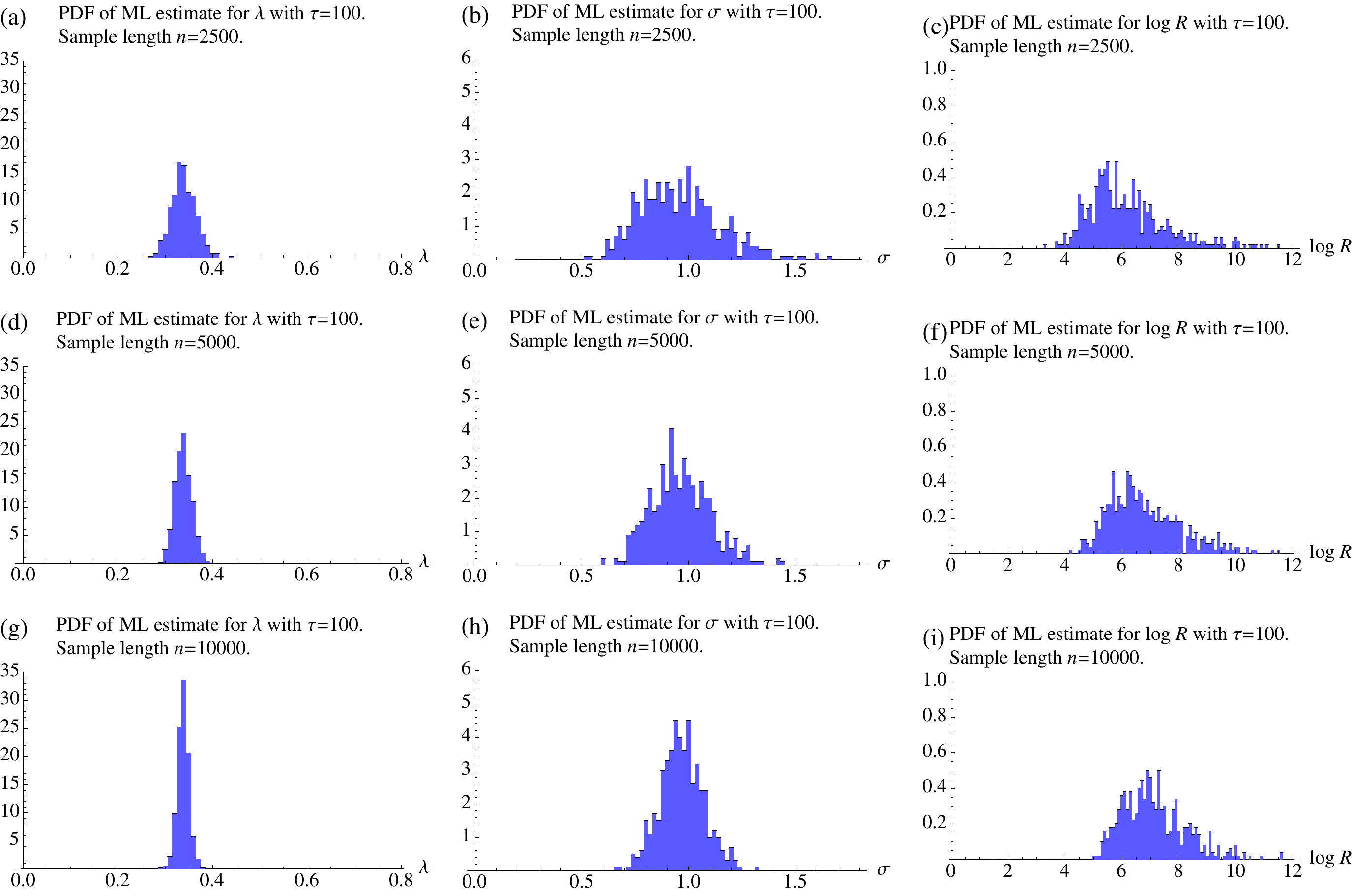}
\caption{The results of the Monte Carlo study for the ML estimator with $\tau=100$. The figures show the estimated probability density functions for the estimators based on 500 realizations of the process. The parameters are $\lambda=0.35$, $\sigma=1$ and $R=2000$ (i.e. $\log R = 7.6$). In figures (a-c) the sample lengths are $n=2500$, in figures (d-f) the sample lengths are $n=5000$ and in figures (g-i) the sample lengths are $n=10000$. The means and standard deviations of the estimators are reported in table \ref{tab2}.} \label{Fig3}
\end{center}
\end{figure*}
%**********************************
%%
%%

%**********************************
\noindent
\begin{figure}%[t]
\begin{center}
\includegraphics[width=6.9cm]{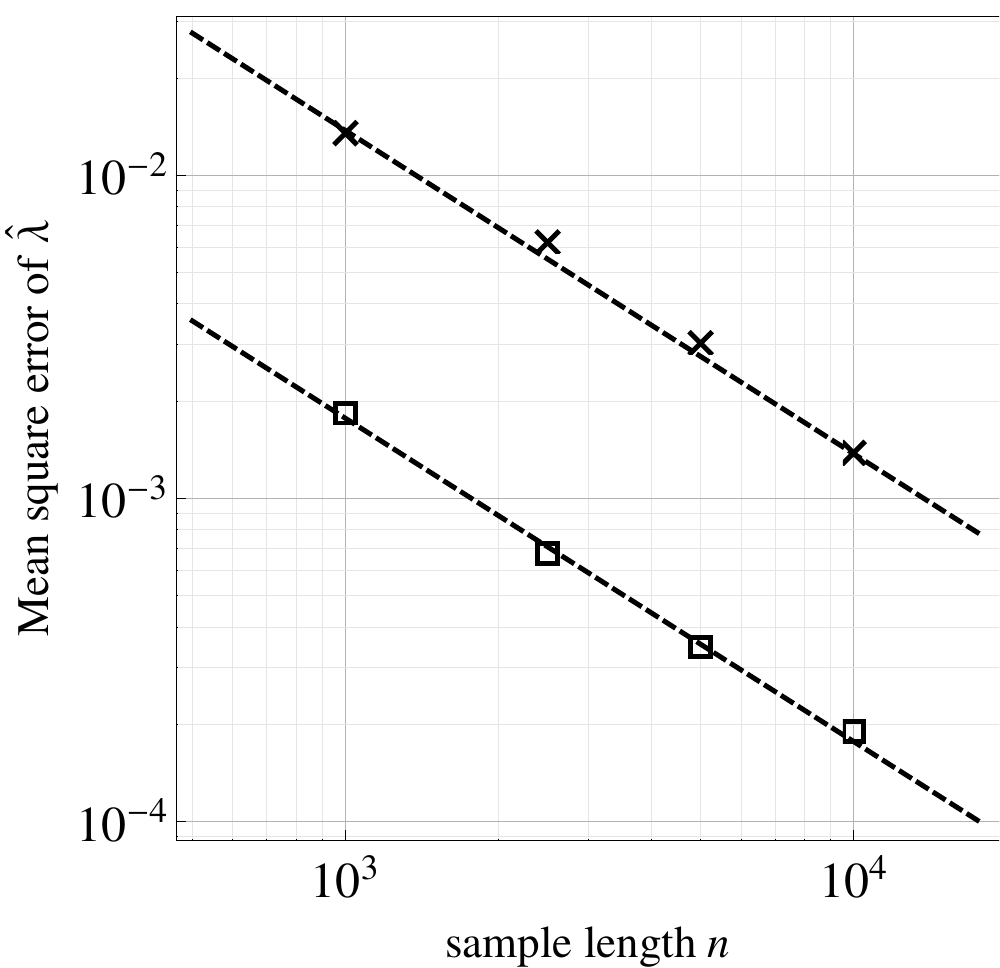}
\caption{Double-logarithmic plot of the mean square errors as functions of sample length $n$ for the ML estimator with $\tau=100$ (squares) and the GMM estimator (crosses). The dotted lines have slopes equal to $-1$, i.e. the mean square errors decay roughly as $1/n$ for both estimators.} \label{Fig4}
\end{center}
\end{figure}
%**********************************
%%
%%

\section{Estimator comparisons} \label{montecarlo}
In this section the ML estimator is compared with an GMM approach which is similar to the one used in \cite{Bacry2008}. This GMM version is essentially a least-square fitting of the auto-correlation function for the logarithmic volatility, and we briefly explain this method in the following:   
Denote $m_t = \log x_t^2$ and observe that 
$$
m_t = h_t + y_t 
$$
where $y_t = \log c + \log \varepsilon_t^2$ are independent and identically distributed. 
We can use the sample standard deviation to normalize $m_t$ so that it has unit variance. Then, 
if we let $\mu_m = \E[m_t] = \E[y_t]$ denote the mean of $m_t$, the auto-correlation function of $m_t$ has the form 
\begin{equation*}
\begin{array}{lll}
\op{ACF}_m(t) &=& \displaystyle \E[(m_1-\mu_m)(m_{t+1}-\mu_m)] \\ &=& \displaystyle  \E[h_1 h_{t+1}] = \lambda^2  \log^+ \frac{R}{t+1}\,.
\end{array}
\end{equation*}
For $t \leq R$ we have 
$$
\op{ACF}_m(t) = \lambda^2 \log R - \lambda^2 \log (t+1)\,,
$$
and $\log R$ and $\lambda$ can be found by linear regression of the auto-correlation function versus $\log (t+1)$.  
\color{black}

We begin testing the approximated ML estimator by applying it to various stock market indices. We use daily log-returns and in all of the estimates the truncation parameter is set to $\tau=500$ days. The results are presented in table \ref{tab1}. We observe that the intermittency parameter $\lambda$ varies from $0.29$ to $0.37$ for the different indices and time periods. We also observe that the  correlation range parameter $T$ varies by roughly one order of magnitude, in the range 1.4-12.2 years. If we compare with the GMM we see that, for all the indices, the estimates of $\lambda$ are lower using the ML method.  For the parameter $T$ the estimates using ML and GMM are more or less consistent, but with quite large variations between the two estimators.  

To further test the performance of the proposed ML estimator we run a small-sample Monte Carlo study. We have used three different sample lengths $n\in \{2500, 5000, 10000\}$, and for each sample length $n$ we simulated 500 sample realizations. The parameter vector considered is $\lambda=0.35$, $\sigma=1$ and $R=2000$.
For the truncation parameter we have considered the cases $\tau\in\{10, 50, 100\}$, and in the GMM method we use a maximum time lag $t_{\op{max}} =500$ days in the auto-correlation function of $m_t=\log x_t^2$.

The results are presented in table \ref{tab2}. For both the GMM and the ML methods the estimates of $R$ are highly unstable. This is also pointed out in \cite{Bacry2008}. 
However, the processes $x_t$ only depend on the $R$ through expressions on the form $\lambda^2 \log R$. Therefore, in order to have an estimator which is comparable to $\lambda$, we should consider the variable $\log R$. The estimators of $\log R$ behave reasonably well, even though there are significant mean square errors and some bias. We see that both the ML and GMM method underestimate $\log R$ and that the errors are roughly the same for the two estimators. 

On the other hand we observe that the ML estimates of $\lambda$ have a standard deviations which are much smaller than the corresponding standard deviation for the GMM estimate, especially for $\tau=100$.  
This can also be seen from figures \ref{Fig2} and \ref{Fig3} where the probability density functions for the GMM estimates and the ML estimates are presented. Based on this we conclude that the ML estimator performs better than the GMM. Moreover, if one allows longer computing times, the truncation parameter $\tau$ can be increased to obtain even more accurate estimates. For a time series of $n=10^4$ data points, an ML estimate with $\tau=500$ takes a few minutes on a personal computer.

In figure \ref{Fig4} we have plotted the mean square errors (MSE) $\mathbb{E}[(\hat{\lambda}-\lambda)^2]$ for the ML estimator with $\tau=100$ and the GMM estimator. We see that for both the estimators the MSE is roughly invserly proportional to the sample length. However, from table \ref{tab2} we see that is a slight negative bias in $\hat{\lambda}$ for the ML estimator. This bias decreases with increasing $\tau$, and we suspect the estimator to be asymptotically unbiased in the limit $\tau=n \to \infty$.

\section{Concluding remarks}
In this paper we have presented an approximate ML estimator for MRW processes. The method is implemented and tested in a Monte Carlo study, and the results show significant improvements over existing methods for the intermittency parameter $\lambda$. 

The methods of this paper represent a suitable starting point for two important generalizations. The first generalization is to allow for correlated innovations, for instance by letting $\varepsilon_t$ be a fractional Gaussian noise. This has several important applications, for instance in modeling of geomagnetic activity \cite{Rypdal2010,Rypdal2011} and electricity spot prices \cite{Malo2006}. Another interesting generalization is to consider the non-Gaussian IDC models referred to in section \ref{model}.  

We also point out that the techniques presented in section \ref{method} can be used to calculate conditional densities on the form 
$p(x_{t+1},\dots,x_{t+s}\,|\,x_1,\dots,x_t)$. At time $t$, such an expression provides a complete forecast over the next $s$ time steps. 
Forecasting and risk analysis based on the MRW model and the methods in this paper is a promising topic that will be pursued in future work. \\

\noindent {\bf Acknowledgment.}
This project was supported by {\em Sparebank 1 Nord-Norge} and the Norwegian Research Council (project number 208125).    
We thank K. Rypdal and the anonymous referee for useful comments and suggestions.


\newpage

\begin{thebibliography}{33}
\expandafter\ifx\csname natexlab\endcsname\relax\def\natexlab#1{#1}\fi
\expandafter\ifx\csname bibnamefont\endcsname\relax
  \def\bibnamefont#1{#1}\fi
\expandafter\ifx\csname bibfnamefont\endcsname\relax
  \def\bibfnamefont#1{#1}\fi
\expandafter\ifx\csname citenamefont\endcsname\relax
  \def\citenamefont#1{#1}\fi
\expandafter\ifx\csname url\endcsname\relax
  \def\url#1{\texttt{#1}}\fi
\expandafter\ifx\csname urlprefix\endcsname\relax\def\urlprefix{URL }\fi
\providecommand{\bibinfo}[2]{#2}
\providecommand{\eprint}[2][]{\url{#2}}

\bibitem[{\citenamefont{Obukhov}(1962)}]{Obukhov1962}
\bibinfo{author}{\bibfnamefont{A.~M.} \bibnamefont{Obukhov}},
  \bibinfo{journal}{J. Geophys. Res.} \textbf{\bibinfo{volume}{67}},
  \bibinfo{pages}{3011} (\bibinfo{year}{1962}).

\bibitem[{\citenamefont{Kolmogorov}(1962)}]{Kolmogorov1962}
\bibinfo{author}{\bibfnamefont{A.~N.} \bibnamefont{Kolmogorov}},
  \bibinfo{journal}{J. Fluid. Mech.} \textbf{\bibinfo{volume}{13}},
  \bibinfo{pages}{83} (\bibinfo{year}{1962}).

\bibitem[{\citenamefont{Kolmogorov}(1941)}]{Kolmogorov1941}
\bibinfo{author}{\bibfnamefont{A.~N.} \bibnamefont{Kolmogorov}},
  \bibinfo{journal}{Dokl. Akad. Nauk. SSSR} \textbf{\bibinfo{volume}{31}},
  \bibinfo{pages}{301} (\bibinfo{year}{1941}).

\bibitem[{\citenamefont{Kahane}(1985)}]{Kahane1985}
\bibinfo{author}{\bibfnamefont{J.~P.} \bibnamefont{Kahane}},
  \bibinfo{journal}{Ann. Sci. Math. Quebec} \textbf{\bibinfo{volume}{9}},
  \bibinfo{pages}{105} (\bibinfo{year}{1985}).

\bibitem[{\citenamefont{Riedi and Ribeiro}(1999)}]{Riedi1999}
\bibinfo{author}{\bibfnamefont{R.~H.} \bibnamefont{Riedi}} \bibnamefont{and}
  \bibinfo{author}{\bibfnamefont{V.~J.} \bibnamefont{Ribeiro}},
  \bibinfo{journal}{IEEE T. Inform. Theory} \textbf{\bibinfo{volume}{45}},
  \bibinfo{pages}{992} (\bibinfo{year}{1999}).

\bibitem[{\citenamefont{Rypdal and Rypdal}(2010)}]{Rypdal2010}
\bibinfo{author}{\bibfnamefont{M.}~\bibnamefont{Rypdal}} \bibnamefont{and}
  \bibinfo{author}{\bibfnamefont{K.}~\bibnamefont{Rypdal}},
  \bibinfo{journal}{J. Geophys. Res.} \textbf{\bibinfo{volume}{115}},
  \bibinfo{pages}{A11216} (\bibinfo{year}{2010}).

\bibitem[{\citenamefont{Rypdal and Rypdal}(2011)}]{Rypdal2011}
\bibinfo{author}{\bibfnamefont{M.}~\bibnamefont{Rypdal}} \bibnamefont{and}
  \bibinfo{author}{\bibfnamefont{K.}~\bibnamefont{Rypdal}},
  \bibinfo{journal}{J. Geophys. Res.} \textbf{\bibinfo{volume}{116}},
  \bibinfo{pages}{A02202} (\bibinfo{year}{2011}).

\bibitem[{\citenamefont{Pathirana and Herath}(2002)}]{Pathirana2002}
\bibinfo{author}{\bibfnamefont{A.}~\bibnamefont{Pathirana}} \bibnamefont{and}
  \bibinfo{author}{\bibfnamefont{S.}~\bibnamefont{Herath}},
  \bibinfo{journal}{Hydrol. Earth Syst. Sc.} \textbf{\bibinfo{volume}{6}},
  \bibinfo{pages}{695} (\bibinfo{year}{2002}).

\bibitem[{\citenamefont{Ghashghaie et~al.}(1996)\citenamefont{Ghashghaie,
  Brewmann, Peinke, Talkner, and Dodge}}]{Ghashghaie1996}
\bibinfo{author}{\bibfnamefont{S.}~\bibnamefont{Ghashghaie}},
  \bibinfo{author}{\bibfnamefont{S.}~\bibnamefont{Brewmann}},
  \bibinfo{author}{\bibfnamefont{W.}~\bibnamefont{Peinke}},
  \bibinfo{author}{\bibfnamefont{J.}~\bibnamefont{Talkner}}, \bibnamefont{and}
  \bibinfo{author}{\bibfnamefont{Y.}~\bibnamefont{Dodge}},
  \bibinfo{journal}{Nature} \textbf{\bibinfo{volume}{381}},
  \bibinfo{pages}{767} (\bibinfo{year}{1996}).

\bibitem[{\citenamefont{Mandelbrot et~al.}(1997)\citenamefont{Mandelbrot,
  Fisher, and Calvet}}]{Mandelbrot1997}
\bibinfo{author}{\bibfnamefont{B.}~\bibnamefont{Mandelbrot}},
  \bibinfo{author}{\bibfnamefont{A.}~\bibnamefont{Fisher}}, \bibnamefont{and}
  \bibinfo{author}{\bibfnamefont{L.}~\bibnamefont{Calvet}},
  \bibinfo{journal}{Cowles Foundation for Research in Economics Working Papers}
   (\bibinfo{year}{1997}).

\bibitem[{\citenamefont{Di~Matteo}(2007)}]{DiMatteo2007}
\bibinfo{author}{\bibfnamefont{T.}~\bibnamefont{Di~Matteo}},
  \bibinfo{journal}{Quant. Finance} \textbf{\bibinfo{volume}{7}},
  \bibinfo{pages}{21} (\bibinfo{year}{2007}).

\bibitem[{\citenamefont{Calvet and Fisher}(2001)}]{Calvet2001}
\bibinfo{author}{\bibfnamefont{L.}~\bibnamefont{Calvet}} \bibnamefont{and}
  \bibinfo{author}{\bibfnamefont{A.}~\bibnamefont{Fisher}},
  \bibinfo{journal}{J. Econometrics} \textbf{\bibinfo{volume}{105}},
  \bibinfo{pages}{27 } (\bibinfo{year}{2001}).

\bibitem[{\citenamefont{Mandelbrot}(1963)}]{Mandelbrot1963}
\bibinfo{author}{\bibfnamefont{B.~B.} \bibnamefont{Mandelbrot}},
  \bibinfo{journal}{J. Bus.} \textbf{\bibinfo{volume}{36}},
  \bibinfo{pages}{394} (\bibinfo{year}{1963}).

\bibitem[{\citenamefont{Lux}(2003)}]{Lux2003}
\bibinfo{author}{\bibfnamefont{T.}~\bibnamefont{Lux}}, \bibinfo{type}{Economics
  Working Papers}, \bibinfo{institution}{Christian-Albrechts-University of
  Kiel} (\bibinfo{year}{2003}).

\bibitem[{\citenamefont{Chapman et~al.}(2005)\citenamefont{Chapman, Hnat,
  Rowlands, and Watkins}}]{Chapman2005}
\bibinfo{author}{\bibfnamefont{S.~C.} \bibnamefont{Chapman}},
  \bibinfo{author}{\bibfnamefont{B.}~\bibnamefont{Hnat}},
  \bibinfo{author}{\bibfnamefont{G.}~\bibnamefont{Rowlands}}, \bibnamefont{and}
  \bibinfo{author}{\bibfnamefont{N.~W.} \bibnamefont{Watkins}},
  \bibinfo{journal}{Nonlinear Proc. Geoph.} \textbf{\bibinfo{volume}{12}},
  \bibinfo{pages}{767} (\bibinfo{year}{2005}).

\bibitem[{\citenamefont{Lux}(2008)}]{Lux2008}
\bibinfo{author}{\bibfnamefont{T.}~\bibnamefont{Lux}}, \bibinfo{journal}{J.
  Bus. Econ. Stat.} \textbf{\bibinfo{volume}{26}}, \bibinfo{pages}{194}
  (\bibinfo{year}{2008}).

\bibitem[{\citenamefont{Bacry et~al.}(2001)\citenamefont{Bacry, Delour, and
  Muzy}}]{Bacry2001}
\bibinfo{author}{\bibfnamefont{E.}~\bibnamefont{Bacry}},
  \bibinfo{author}{\bibfnamefont{J.}~\bibnamefont{Delour}}, \bibnamefont{and}
  \bibinfo{author}{\bibfnamefont{J.~F.} \bibnamefont{Muzy}},
  \bibinfo{journal}{Phys. Rev. E} \textbf{\bibinfo{volume}{64}},
  \bibinfo{pages}{026103} (\bibinfo{year}{2001}).

\bibitem[{Note1()}]{Note1}
Note1, \bibinfo{note}{in turbulence $T$ corresponds to the integral scale.}

\bibitem[{\citenamefont{Bacry et~al.}(2008)\citenamefont{Bacry, Kozhemyak, and
  Muzy}}]{Bacry2008}
\bibinfo{author}{\bibfnamefont{E.}~\bibnamefont{Bacry}},
  \bibinfo{author}{\bibfnamefont{A.}~\bibnamefont{Kozhemyak}},
  \bibnamefont{and} \bibinfo{author}{\bibfnamefont{J.-F.} \bibnamefont{Muzy}},
  \bibinfo{journal}{J. Econ. Dyn. Control} \textbf{\bibinfo{volume}{32}},
  \bibinfo{pages}{156} (\bibinfo{year}{2008}).

\bibitem[{\citenamefont{{R Development Core Team}}(2011)}]{R}
\bibinfo{author}{\bibnamefont{{R Development Core Team}}}
  (\bibinfo{year}{2011}), \urlprefix\url{http://www.R-project.org/}.

\bibitem[{\citenamefont{L{\o}vsletten and Rypdal}(2011)}]{code}
\bibinfo{author}{\bibfnamefont{O.}~\bibnamefont{L{\o}vsletten}}
  \bibnamefont{and} \bibinfo{author}{\bibfnamefont{M.}~\bibnamefont{Rypdal}},
  \bibinfo{type}{R package} (\bibinfo{year}{2011}),
  \urlprefix\url{http://complexityandplasmas.net/nordforsk/Papers_files/MLE%20MRW%20R%20package.zip}.

\bibitem[{\citenamefont{Robert and Vargas}(2010)}]{Robert2010}
\bibinfo{author}{\bibfnamefont{R.}~\bibnamefont{Robert}} \bibnamefont{and}
  \bibinfo{author}{\bibfnamefont{V.}~\bibnamefont{Vargas}},
  \bibinfo{journal}{Ann. Probab.} \textbf{\bibinfo{volume}{38}},
  \bibinfo{pages}{605} (\bibinfo{year}{2010}).

\bibitem[{\citenamefont{Bacry and Muzy}(2003)}]{Bacry2003}
\bibinfo{author}{\bibfnamefont{E.}~\bibnamefont{Bacry}} \bibnamefont{and}
  \bibinfo{author}{\bibfnamefont{J.~F.} \bibnamefont{Muzy}},
  \bibinfo{journal}{Commun. Math. Phys.} \textbf{\bibinfo{volume}{236}},
  \bibinfo{pages}{449} (\bibinfo{year}{2003}).

\bibitem[{\citenamefont{Muzy and Bacry}(2002)}]{Muzy2002}
\bibinfo{author}{\bibfnamefont{J.~F.} \bibnamefont{Muzy}} \bibnamefont{and}
  \bibinfo{author}{\bibfnamefont{E.}~\bibnamefont{Bacry}},
  \bibinfo{journal}{Phys. Rev. E} \textbf{\bibinfo{volume}{66}},
  \bibinfo{pages}{056121} (\bibinfo{year}{2002}).
  
  

  \bibitem[{\citenamefont{McLeod et~al.}(2007)}]{McLeod2007}
\bibinfo{author}{\bibfnamefont{I.~A.} \bibnamefont{McLeod}},
 \bibinfo{author}{\bibfnamefont{H.}~\bibnamefont{Yu}}, \bibnamefont{and} 
  \bibinfo{author}{\bibfnamefont{Z.~L.}~\bibnamefont{Krougly}}, 
  \bibinfo{journal}{J. Stat. Softw.} \textbf{\bibinfo{volume}{23}},
  \bibinfo{pages}{1} 
 (\bibinfo{year}{2007}).


\bibitem[{\citenamefont{Levinson}(1946)}]{Levinson1946}
\bibinfo{author}{\bibfnamefont{N.}~\bibnamefont{Levinson}},
  \bibinfo{journal}{J. Math. Phys.} \textbf{\bibinfo{volume}{25}},
  \bibinfo{pages}{261} (\bibinfo{year}{1946}).

\bibitem[{\citenamefont{Trench}(1964)}]{Trench1964}
\bibinfo{author}{\bibfnamefont{W.~F.} \bibnamefont{Trench}},
  \bibinfo{journal}{J. Soc. Indust. Appl. Math.} \textbf{\bibinfo{volume}{12}},
  \bibinfo{pages}{515} (\bibinfo{year}{1964}).

\bibitem[{\citenamefont{Skaug and Yu}(2009)}]{Skaug2009}
\bibinfo{author}{\bibfnamefont{H.}~\bibnamefont{Skaug}} \bibnamefont{and}
  \bibinfo{author}{\bibfnamefont{J.}~\bibnamefont{Yu}},
  \bibinfo{journal}{Research Collection School of Economics, Singapore
  Management University}  (\bibinfo{year}{2009}).

\bibitem[{\citenamefont{Martino et~al.}(2011)\citenamefont{Martino, Aas,
  Lindqvist, Neef, and Rue}}]{Aas2011}
\bibinfo{author}{\bibfnamefont{S.}~\bibnamefont{Martino}},
  \bibinfo{author}{\bibfnamefont{K.}~\bibnamefont{Aas}},
  \bibinfo{author}{\bibfnamefont{O.}~\bibnamefont{Lindqvist}},
  \bibinfo{author}{\bibfnamefont{L.~R.} \bibnamefont{Neef}}, \bibnamefont{and}
  \bibinfo{author}{\bibfnamefont{H.}~\bibnamefont{Rue}},
  \bibinfo{journal}{Europ. J. Finance} \textbf{\bibinfo{volume}{17}},
  \bibinfo{pages}{487} (\bibinfo{year}{2011}).

\bibitem[{\citenamefont{Cruz et~al.}(2006)\citenamefont{Cruz, Mart{\'\i}nez,
  and Raydan}}]{Raydan2006}
\bibinfo{author}{\bibfnamefont{W.~L.} \bibnamefont{Cruz}},
  \bibinfo{author}{\bibfnamefont{J.}~\bibnamefont{Mart{\'\i}nez}},
  \bibnamefont{and} \bibinfo{author}{\bibfnamefont{M.}~\bibnamefont{Raydan}},
  \bibinfo{journal}{Math. Comput.} \textbf{\bibinfo{volume}{75}},
  \bibinfo{pages}{1429} (\bibinfo{year}{2006}).

\bibitem[{\citenamefont{Varadhan and Gilbert}(2009)}]{BB}
\bibinfo{author}{\bibfnamefont{R.}~\bibnamefont{Varadhan}} \bibnamefont{and}
  \bibinfo{author}{\bibfnamefont{P.}~\bibnamefont{Gilbert}},
  \bibinfo{journal}{J. Stat. Softw.} \textbf{\bibinfo{volume}{32}},
  \bibinfo{pages}{1} (\bibinfo{year}{2009}).

\bibitem[{\citenamefont{Bates and Maechler}(2011)}]{Matrix}
\bibinfo{author}{\bibfnamefont{D.}~\bibnamefont{Bates}} \bibnamefont{and}
  \bibinfo{author}{\bibfnamefont{M.}~\bibnamefont{Maechler}}, \bibinfo{type}{R
  package} (\bibinfo{year}{2011}),
  \urlprefix\url{http://CRAN.R-project.org/package=Matrix}.

\bibitem[{yah(~)}]{yahoo}
%\bibinfo{type}{Tech. Rep.} (\bibinfo{year}{~}),
  \urlprefix\url{http://finance.yahoo.com/}.

\bibitem[{\citenamefont{Malo}(2006)}]{Malo2006}
\bibinfo{author}{\bibfnamefont{P.}~\bibnamefont{Malo}},
  \bibinfo{journal}{Helsinki School of Economics Working Papers}
  (\bibinfo{year}{2006}).

\end{thebibliography}
\end{document}